\documentclass[superscriptaddress,groupedaddress,nofootnoteinbib,12pt]{article}
\pdfoutput=1
\usepackage{color}
\usepackage{graphicx}
\usepackage{dcolumn}
\usepackage{bm}
\usepackage{amssymb}
\usepackage{amsmath}
\usepackage{sectsty}
\usepackage{colortbl}
\usepackage{latexsym}
\usepackage{float}
\usepackage{ifthen}
\usepackage{enumerate}
\usepackage{url}
\usepackage[force]{feynmp-auto}
\usepackage{jcappub}
    \usepackage{picinpar}
    \usepackage{colortbl}
\usepackage{multirow}
	\usepackage{float}
	      \usepackage{setspace}
\usepackage{array}
\usepackage{bm}
\usepackage{amsopn}

\usepackage{booktabs}
\usepackage[table]{xcolor}
\definecolor{lightgray}{gray}{0.9}

\newcommand{\ie}{{\it i.~e.}}

\def\d{\textrm{d}}

\newcommand{\p}{\partial}

\def\ba{\begin{eqnarray}}
\def\ea{\end{eqnarray}}
\def\beq{\begin{eqnarray}}
\def\eeq{\end{eqnarray}}

\def\mpl{M_{\rm Pl}}

\def\L{\mathcal{L}}

\def\({\left(}
\def\){\right)}
\def\ie{{\it i.e.}}

\def\nn{\nonumber}
\def\mn{_{\mu \nu}}
\def\stu{St\"uckelberg }
\def\p{\partial}
\def\mupn{^\mu_{\ \nu}}
\def\<{\langle}
\def\>{\rangle}

\def\K{\mathcal{K}}

\def\gd{g_{\rm eff}}
\def\gu{g^{\rm eff}}

\usepackage{titlesec}
\def\stu{St\"uckelberg }

\newcommand{\subpara}[1]{\par\vspace{1mm}\noindent\emph{{#1}}.---}

\newcolumntype{Q}{>{$\displaystyle}l<{$}}
\newcolumntype{q}{>{\columncolor[gray]{0.9}$\displaystyle}l<{$}}
\newcolumntype{R}{>{$\displaystyle}r<{$}}
\newcolumntype{S}{>{$\displaystyle}c<{$}}
\newcolumntype{s}{>{\columncolor[gray]{0.9}$\displaystyle}c<{$}}
\newcolumntype{T}{>{\columncolor[gray]{0.9}}c<{}}

\newsavebox{\tableA}
\newsavebox{\tableB}

\newsavebox{\boxplot}
\newsavebox{\boxplota}

\definecolor{dullpurple}{rgb}{0.431,0.188,0.534}
\definecolor{darkgreen}{rgb}{0.133,0.545,0.133}

\newboolean{editorial}
\setboolean{editorial}{true}
\newcommand{\editorial}[2]{\ifthenelse{\boolean{editorial}}{\textcolor{red}{ [\textsf{\textbf{{#1}}}: } \textcolor{blue}{\textsf{{#2}}]}}{}\\}

\begin{document}

\title{
On couplings to matter in massive (bi-)gravity
}
	
\author{Claudia de Rham$^{a,b}$, Lavinia Heisenberg$^{b,c}$ and Raquel H. Ribeiro$^{a}$}
\affiliation{$^{a}$CERCA/Department of Physics, Case Western Reserve University, \\
10900 Euclid Ave, Cleveland, OH 44106, U.S.A.}
\affiliation{$^{b}$Perimeter Institute for Theoretical Physics, \\
31 Caroline St N, Waterloo, Ontario, N2L 6B9, Canada}
\affiliation{$^{c}$D\'epartment de Physique  Th\'eorique and Center for
Astroparticle Physics, \\ Universit\'e de  Gen\`eve, 24 Quai E. Ansermet, CH-1211  Gen\`eve, Switzerland}

	\emailAdd{Claudia.deRham@case.edu}
    \emailAdd{Lavinia.Heisenberg@unige.ch}
	\emailAdd{RaquelHRibeiro@case.edu}

\abstract{
We investigate the coupling to matter in ghost-free massive (bi-)gravity. When species in the matter sector couple covariantly to only one metric, we show that at one--loop these couplings do not spoil the special structure of the graviton potential. When the same species couples directly to both metrics we show that a ghost is present at the classical level and that loops destroy the special structure of the potential at an unacceptably low scale. We then propose a new `composite' effective metric built out of both metrics.
When matter fields couple covariantly to this effective metric,  the would be Boulware--Deser ghost is absent
in different representative limits. At one--loop such couplings do not detune the special structure of the potential.  We conjecture that matter can couple covariantly to that effective metric in all generality without introducing any Boulware--Deser ghost below a cut-off scale parametrically larger than the strong coupling scale. We also discuss alternative couplings to matter where the kinetic and potential terms of the matter field couple to different metrics. In both cases we discuss preliminary implications for cosmology.
}	

\maketitle

\section{Introduction}
\label{sec:introduction}

The recent years have seen an increased interest in theories of modified gravity in the infrared and their connections with the late behaviour of the Universe, see Ref.~\cite{Joyce:2014kja} for a recent review.
Despite many advances in our understanding of massive gravity \cite{deRham:2010ik,deRham:2010kj}, bi-gravity \cite{Hassan:2011zd} and multi-gravity \cite{Hinterbichler:2012cn} (see Refs.~\cite{Hinterbichler:2011tt,deRham:2014zqa,Comelli:2014xga}), much remains to be explored in these types of theories including their stability under quantum effects and their couplings to matter. In the same way General Relativity (GR) is consistently constructed so as to be compatible with the coupling to the matter sector, the theories of massive (bi-)gravity should satisfy the same criterion. \\

A first  worry in these types of theories is the existence of superluminal propagation as has been established for the  fluctuations of the \stu fields about Vainshtein-like configurations in DGP \cite{Dvali:2000hr,Hinterbichler:2009kq} and  massive gravity \cite{Burrage:2011cr,Deser:2012qx,Izumi:2013poa,Deser:2013eua,Deser:2013qza,Deser:2014hga} and bi- or multi-gravity. More generically, theories which behave as a  Galileon \cite{Nicolis:2008in} or DBI--Galileon \cite{deRham:2010eu} in some limit can propagate a classical superluminally group velocity {\it within} in the regime of the effective field theory \cite{Nicolis:2008in,Goon:2010xh,deFromont:2013iwa}. However, causality and the production of closed-timelike curves is not determined by the group velocity but rather by the front velocity (large frequency limit of the phase velocity \cite{Hollowood:2007kt,Shore:2007um}) which cannot be determined within the classical regime of validity of the theory. As shown in Ref.~\cite{Burrage:2011cr} for Galileons and within the decoupling of massive gravity,  closed-timelike curves cannot be produced within the regime of validity of the theory.
Parallel to these studies, a dual description to these types of theory has been developed in Refs.~\cite{Curtright:2012gx,Creminelli:2013fxa,deRham:2013hsa,Creminelli:2014zxa, deRham:2014lqa, Kampf:2014rka} in which a theory that admits  superluminal group velocity can be  mapped into a (sub)luminal theory  with analytic S-matrix \cite{deRham:2013hsa}. Such a map comes to show how subtle these issues are and how well-behaved theories may still exhibit superluminal group velocities. This could also open the door to a better understanding of these types of theories (which exhibit the Vainshtein mechanism \cite{Vainshtein:1971ip}) at the quantum level. \\

Besides the previous considerations, the theoretical viability of massive gravity and its bi- and multi-gravity extensions relies on a very precise form of the interactions or mass term \cite{deRham:2010ik} which generalises the linear Fierz--Pauli structure \cite{Fierz:1939ix} and prevents the presence of Boulware--Deser (BD) ghost \cite{Boulware:1973my}. On the other hand, the observational viability of these models relies on a very small mass parameter, typically of the order of the Hubble parameter today, together with a successful implementation of the Vainshtein mechanism \cite{Vainshtein:1971ip}.\\

These requirements can be satisfied at the classical level but quantum effects could {\it a priori} lead to large corrections to the graviton mass  or destabilise the delicate ghost-free structure of the potential interactions, potentially bringing back the BD ghost at unacceptably low scales. Within the context of massive gravity, these effects were studied in Refs.~\cite{deRham:2012ew,deRham:2013qqa}. In Ref.~\cite{deRham:2012ew} it was shown that the graviton mass is radiatively stable against quantum corrections and thus technically natural. Setting it to arbitrarily small values thus represents a tuning but one which does not get spoiled by quantum corrections. \\

Still within the context of massive gravity, the effects of quantum corrections  on the special structure of the graviton potential were explored in Ref.~\cite{deRham:2013qqa}. While graviton loops typically destabilise the structure of the graviton potential, they do so at a scale which is harmless and effectively inconsequential for the theory. An analogous effect is found in scalar theories which rely on derivative interactions, such as $P(X)$ and galileon models \cite{Nicolis:2008in}: in some limits these theories are also stable against quantum effects within the regime of validity of their effective field theory \cite{Nicolis:2004qq, dePaulaNetto:2012hm, Brouzakis:2013lla, Brouzakis:2014bwa, deRham:2014wfa, Heisenberg:2014raa}.\\

 As for the matter loops, the contributions to the graviton potential are identical to those in GR as long as matter couples covariantly to the dynamical metric. The resulting quantum corrections are thus nothing other than the standard cosmological constant which does not affect the special structure of the graviton potential and is thus harmless (from what concerns the BD ghost\footnote{Massive gravity does not prevent the vacuum energy from acquiring large quantum corrections. If massive gravity was ever to tackle the old cosmological constant problem it would more likely be via the idea of degravitation \cite{Dvali:2002fz,Dvali:2002pe,ArkaniHamed:2002fu,Dvali:2007kt,deRham:2010tw} where the vacuum energy or cosmological constant could be large but have a small effect on the geometry. See also Ref.~\cite{Gabadadze:2014rwa} for recent developments.}). This result is maintained at all loops and in this work we shall show that it trivially generalises to bi-gravity, as long as the matter field only couples directly to one metric. \\

Within the context of bi- and multi-gravity, more general couplings to matter have been explored in the literature, some of which can lead to an interesting new phenomenology \cite{Khosravi:2011zi,Akrami:2012vf,Akrami:2013ffa,Akrami:2014lja,deRham:2013awa} see also \cite{Tamanini:2013xia} recent work directly in the vielbein formalism. In  bi-gravity, both metrics are put on an equal footing, and coupling matter to both metrics simultaneously might thus appear natural at first sight. In this manuscript we shall see that the same matter field cannot have two kinetic terms (one with respect to each metric) without reintroducing the BD ghost. We show that using three different methods, two classical and one quantum. Whilst these results were derived explicitly for bi-gravity it is straight-forward to show that they hold for any multi-gravity theory where the matter field couples covariantly to more than one metric.\\

To prove these results we show in the decoupling limit and in the mini-superspace approximation that if a matter field couples covariantly to both metrics in bi-gravity, the BD ghost re-emerges at an unacceptable low-scale, ruling out any predictions of the theory. We also show that the quantum corrections from the matter field would destabilise the structure of the potential, and this also at an unacceptable low scale. As a result, while bi-gravity may enjoy a one-to-one correspondence between its two metrics, this correspondence is usually broken by the matter sector for a consistent theory. As we present in this manuscript, there may be exceptional cases where the matter sector couples to an effective metric for which the one-to-one correspondence between the two metrics is maintained. However even in the exceptional case presented here, a ghost is expected to appear below the Planck scale. It is then unclear whether the new physics that ought to enter at or below the mass of the would-be-ghost can maintain the one-to-one correspondence between the two metrics. This would be interesting to understand more carefully.  \\

In the rest of the manuscript, we explore a new type of coupling to matter where the matter fields see an effective metric $\gu$ built out of both metrics $g\mn$ and $f\mn$ in bi-gravity (or built out of both the dynamical metric $g\mn$ and reference metric $f\mn$ in massive gravity),
\ba
\label{eq:geffintro}
\gu\mn=\alpha^2 g\mn +2 \alpha \beta\ g_{\mu \alpha}\, X^\alpha_{\ \nu} +\beta^2 f\mn\,,
\ea
where $\alpha$ and $\beta$ are arbitrary constants. Here and in what follows the matrix $X\mupn$ is defined as $X^\mu_{\, \alpha}X^\alpha_{\, \nu}= g^{\mu\alpha}f_{\alpha\nu}$. Symbolically, we may thus write the matrix $X$ as $X\mupn = (\sqrt{g^{-1}f})^\mu_{\ \nu}$.
When coupled to this effective metric, matter field loops no longer generate a cosmological constant with respect to either metrics but rather a contribution going as $\sqrt{-\det{\gu\mn}}$.
This effective metric was derived by demanding that $\sqrt{-\det{\gu\mn}}$ takes the form of one of the acceptable potentials of massive (bi-)gravity. We then explore this coupling in the mini-superspace and show the absence of ghost in that approximation. This result holds when perturbing about arbitrary FLRW (Friedmann-Lema\^itre-Robertson-Walker) solutions, which implies the existence of non-trivial yet consistent cosmological solutions.
We also explore the coupling of matter to the effective metric in the decoupling-limit of the theory and show that it remains free from the BD ghost. This implies that even if the theory did involve a BD ghost it would do so at a scale above the strong coupling one. The theory can thus be studied as an effective field theory with its own interesting phenomenology below a cut-off scale which is parametrically larger than the strong coupling scale. When performing a proper ADM analysis we find 
a ghost at {\it sixth} order in perturbations about flat space-time which strongly suggests that highly anisotropic solutions of the theory would excite the BD ghost. \\

In massive gravity, the study of cosmological solutions has been notoriously challenging (see Refs.~\cite{deRham:2010tw,deRham:2011by, Comelli:2011zm,Comelli:2012db,Koyama:2011wx,Volkov:2011an,Koyama:2011xz,Chamseddine:2011bu,Gratia:2012wt,Kobayashi:2012fz,Tasinato:2012ze,Wyman:2012iw,Volkov:2012zb,Khosravi:2013axa,Volkov:2013roa,Gratia:2013gka,DeFelice:2013bxa,DeFelice:2013awa,DeFelice:2012mx,Tasinato:2013rza,PhysRevD.88.063006,Motohashi:2012jd,Gratia:2013uza,Volkov:2012cf,Comelli:2013tja, Heisenberg:2014kea}), with more success in its quasi-dilaton and bi-gravity extensions  (see Refs.~\cite{DAmico:2012zv,DAmico:2013kya, Gabadadze:2014kaa,DeFelice:2013tsa,DeFelice:2013dua,Haghani:2013eya} and Refs.~\cite{Fasiello:2013woa,Comelli:2014bqa,Konnig:2013gxa,DeFelice:2013nba,DeFelice:2014nja,Solomon:2014dua,Schmidt-May:2014iea,Konnig:2014xva,Konnig:2014dna,Nojiri:2012zu,Nojiri:2012re,Bamba:2013hza}). The absence of exact FLRW solutions in massive gravity (with flat or closed spatial curvature and with Minkowski reference metric) \cite{DAmico:2011jj} has made the study of cosmological solutions particularly difficult (even though solutions arbitrarily close to FLRW may be found on distances smaller or comparable to the current observable Universe).
Instead when at least one field couples to the effective metric \eqref{eq:geffintro}, the no-go result for the existence of exact FLRW solutions in massive gravity gets broken\footnote{If only one matter field couples to the effective metric then some mild assumptions ought to be imposed on the way that field enters the Lagrangian, however these are relaxed if several species couple to the effective metric. We emphasize however that these fields are expected to belong to a dark sector and we do not consider them to be one of the fields that drive the cosmic expansion of the Universe.} and opens the door to a new set of interesting cosmological solutions without resorting to any new degrees of freedom other than those already present in the massive graviton and the matter fields. We simply discuss this possibility here and leave the study of exact cosmological solutions to further studies.  We emphasize however that since there is no BD ghost for that coupling in the mini-superspace, this ensures that the exact FLRW solutions obtained here are also free of the ghost. Moreover we perform a linear perturbation analysis about these solutions and find no BD ghost. \\

\subpara{Outline}This paper is organised as follows. In section~\ref{sec:review} we review the formalism behind massive (bi-)gravity and the special structure of its potential. We show that when a matter field only couples covariantly to one metric, the quantum corrections do not destabilise the form of the potential. We then investigate the situation where a matter field couples covariantly to two metrics simultaneously in section~\ref{sec:2metricCoupling} and show the emergence of a ghost at an unacceptably low scale.  We then propose a new effective metric in section~\ref{sec:effmet} and discuss its consistency in different limits. Based on that coupling to matter, we  show the existence of FLRW solutions in section~\ref{sec:FRW}. Finally, in section~\ref{sec:OtherCoupling} we consider alternative couplings to matter where the kinetic term and the potential term of the field couple to different metrics. We also discuss the implications for cosmology in that case. Our results are summarises in section~\ref{sec:discussion}.
%

%

\section{Review of massive (bi-)gravity}
\label{sec:review}
Since the seminal work of Fierz \& Pauli~\cite{Fierz:1939ix} there were many attempts to
derive a fully non-linear theory of a massive spin-2 graviton, most of which generate a ghost, now commonly known as the Boulware--Deser ghost~\cite{Boulware:1973my}. In very much the same way as GR (in addition to the Lovelock invariants) is the unique consistent theory for a massless spin-2 graviton, there is a unique family of `allowed potentials' for massive gravity. We briefly review the details of massive (bi-)gravity before discussing its coupling to matter.

\subsection{Massive (bi-)gravity \& allowed potentials}
\label{sec:reviewMG}

The theory of massive gravity with dynamical metric $g\mn$ is given by \cite{deRham:2010ik,deRham:2010kj}
\ba
\label{eq:LagMGR}
\L_{\rm mGR}=\frac{\mpl^2}{2} \sqrt{-g}\( R[g]+\frac{m^2}{4}\sum_{n=0}^4\alpha_n \mathcal{U}_n[\K]\)\,,
\ea
where the potential term are expressed symbolically as $\mathcal{U}_n[\mathcal{K}] = \mathcal{E}  \mathcal{E} \K^n$
in terms of the Levi-Cevita tensors $\mathcal{E}$ and the tensors $\K$ defined as
\ba
\K\mupn[g,f] =\delta\mupn -X\mupn\quad\quad{\rm with}\quad\quad X\mupn \equiv \(\sqrt{g^{-1}f}\)\mupn \,.
\ea
These potentials were shown to take the form of a deformed determinant in Ref.~\cite{Hassan:2011vm}. For instance, we may express the following determinant as the sum of `allowed' potential
\ba
\label{eq:det1}
\det \(\alpha +\beta X\mupn \)=\sum_{n=0}^4 \frac{(-\beta)^n}{n!}(\alpha+\beta)^{4-n} \mathcal{U}_n[\K]\,.
\ea
$\mathcal{U}_0$ corresponds to a cosmological constant and $\mathcal{U}_1$ to the tadpole, while the terms with $n\ge 2$ represent the genuine potential interactions. In massive gravity $f\mn$ is the reference metric. In bi-gravity the reference metric is made dynamical by including a curvature term for $f\mn$
\ba
\L_{\rm bi-gravity}=\L_{\rm mGR}+\frac{M_{{\rm Pl}, f}^2}{2} \sqrt{-f}R[f]\,,
\ea
where the two Planck scales $M_{{\rm Pl}, f}$ and $\mpl$ can differ.

In this formulation, both massive gravity and bi-gravity break one copy of diffeomorphism, which can easily be restored by the introduction of four \stu fields $\phi^a$ (only three of which are dynamical), and by promoting the metric $f\mn$ to its `St\"uckelbergized' version\footnote{In bi-gravity we are completely free to  `St\"uckelbergize' the metric $g\mn$ rather than $f\mn$. See Ref.~\cite{Fasiello:2013woa} for interesting consequences of this choice.} $\tilde f\mn$ defined by
\ba
\tilde f\mn= \p_\mu \phi^a \p_\nu \phi^b f_{ab}\,.
\ea
Unitary gauge is recovered by setting the \stu fields to $\phi^a=x^a$.

\subsection{Covariant coupling to matter}

Massive (bi-)gravity was shown to be free from the BD ghost when matter fields are coupled covariantly to the dynamical metric $g\mn$ \cite{Hassan:2011zd}. One could also easily extend the proof to another sector of matter which couples covariantly to the reference metric $f\mn$. The same holds for bi-gravity where matter fields can couple  covariantly to either metric separately,
\ba
\L_{\rm matter}= \L_{g}[g\mn, \chi_g]+\L_f[f\mn, \chi_ f]\,,
\ea
where the two sectors $\chi_g$ and $\chi_f$ are independent and the couplings are considered to be covariant with respect to the respective metrics $g\mn$ or $f\mn$ (it does not matter whether or not the metric $f\mn$ is dynamical).

At one-loop, the contribution from matter fields to the graviton potential were derived in Ref.~\cite{deRham:2013qqa} for massive gravity and shown to lead to a cosmological constant. This result can be easily understood: if only the matter field propagates in the loops, the loops have no knowledge of the graviton mass or the interactions between both metrics  (or of the \stu fields) and the result is bound to be the same as in GR. The result remains in bi-gravity.
We consider for simplicity two massive scalar fields $\chi_g$ and $\chi_f$ with masses $M_g$ and $M_f$ coupled respectively to $g\mn$ and $f\mn$
\ba
\L_{\rm matter}= -\frac 12 \sqrt{-g} \(g^{\mu\nu}\p_\mu \chi_g \p_\nu \chi_g +M_g^2 \chi_g^2\)
-\frac 12 \sqrt{-f} \(f^{\mu\nu}\p_\mu \chi_f \p_\nu \chi_f +M_f^2 \chi_f^2\)\,.
\ea
Looking at the  one--loop effective action from the matter loops (and recalling that there is no mixing between $\chi_g$ and $\chi_f$ and these fields do not mix at one-loop), we find
\begin{eqnarray}
e^{- S^{({\rm matter-loops})}_{1,\textrm{eff}}( g_{ab}, f_{ab})}&=&\int \mathcal{D}\chi_g  \mathcal{D}\chi_f e^{-\chi_g \left(\frac{\delta^2 S}{\delta \chi_g^2}\right)\chi_g-\chi_f \left(\frac{\delta^2 S}{\delta \phi_f^2}\right)\phi_f}\,.
 \end{eqnarray}
Switching to Euclidean signature this leads to the effective action
\ba
S^{({\rm matter-loops})}_{1,\textrm{eff}}( g_{ab}, f_{ab})&=& \frac 12 \log \det  \left(\frac{\delta^2 S}{\delta \chi_g^2}\right)+\frac 12 \log \det  \left(\frac{\delta^2 S}{\delta \chi_f^2}\right) \\
&=&\frac 12 {\rm Tr} \log \left(\frac{\delta^2 S}{\delta \chi_g^2}\right)+\frac 12 {\rm Tr} \log \left(\frac{\delta^2 S}{\delta \chi_f^2}\right)\\
&=&\frac 12 {\rm Tr} \log \(\sqrt{g}\(g^{\mu\nu} \nabla^{(g)}_\mu \nabla^{(g)}_\nu -M_g^2\)\)\\
&+&\frac 12 {\rm Tr} \log \(\sqrt{f}\(f^{\mu\nu} \nabla^{(f)}_\mu \nabla^{(f)}_\nu -M_f^2\)\) \,,\nn
 \ea
 where $\nabla^{(g)}_\mu$ is the covariant derivative with respect to the metric $g\mn$ and respectively for $\nabla^{(f)}_\mu$.
Evaluating these traces performing dimensional regularization, we find the following running contributions
\ba
\mathcal{L}^{({\rm matter-loops})}_{1, \log }=M_g^4 \sqrt{g}\log(M_g/\mu) +M_f^4\sqrt{f} \log(M_f/\mu) + \text{curvature corrections}\,,
\ea
where $\mu$ is an arbitrary sliding scale, and we only focus on the quantum corrections to the graviton potential. We will thus omit the curvature corrections in the rest of this work.  Exactly in the same way as in GR and in massive gravity (with standard coupling to matter) the contributions to the one loop effective action are pure  cosmological constants and therefore do not detune the special structure of the potential or of the ghost-free interactions between the metrics $f\mn$ and $g\mn$. This conclusion can easily be extended to multi-gravity where  $N$ interacting metrics couple covariantly to their respective matter sector.

\section{Coupling to two metrics}
\label{sec:2metricCoupling}

In this section we consider a single species (a scalar field $\chi_g\equiv\chi_f\equiv \chi$) coupling to both metrics simultaneously
\ba
\L_{\rm matter} &=& {\lambda_g} \L_{g}[g\mn, \chi]+ {\lambda_f} \L_f[f\mn, \chi] \\
&=&-\frac {\lambda_g} 2 \sqrt{-g} \(g^{\mu\nu}\p_\mu \chi \p_\nu \chi +M^2 \chi^2\)
-\frac {\lambda_f}2 \sqrt{-f} \(f^{\mu\nu}\p_\mu \chi \p_\nu \chi +M^2 \chi^2\)\,,
\label{eq:samefieldLag}
\ea
where ${\lambda_g}$ and ${\lambda_f}$ are two (positive) dimensionless parameters which have been introduced for the sole purpose of being able to dial off one of the coupling by setting either ${\lambda_g}$ or ${\lambda_f}$ to zero. 

In what follows we consider the one-loop effective action and find that unlike in the previous case, the resulting contribution to the potential term is no longer a cosmological constant with respect to either metric nor any of the ghost-free potentials. Even more intriguing the resulting ghost from such a potential would come in at an unacceptable low scale. This begs the question of whether or not the ghost only enters quantum mechanically or whether it is already present at the classical level, as we shall see.

\subsection{One-loop detuning of the potential}

The  one--loop effective action from the matter loops of $\chi$ which couples to both metrics, is given by
\ba
S^{({\rm matter-loops})}_{1,\textrm{eff}}( g_{ab}, f_{ab})
&=&\frac 12 {\rm Tr} \log \left(\frac{\delta^2 \(S_g+S_f\)}{\delta \chi_g^2}\right)\\
&=&\frac 12 {\rm Tr} \log \({\lambda_g} \sqrt{g}\(g^{\mu\nu} \nabla^{(g)}_\mu \nabla^{(g)}_\nu -M^2\)+{\lambda_f} \sqrt{f}\(f^{\mu\nu} \nabla^{(f)}_\mu \nabla^{(f)}_\nu -M^2\)\) \nn \\
&=&\frac 12 {\rm Tr} \log \( \sqrt{g_{\rm eff}} \( g^{\mu\nu}_{\rm eff} \nabla^{(g_{\rm eff})}_\mu \nabla^{(g_{\rm eff})}_\nu-M^2 + \cdots\)\)\,,\nn
 \ea
 where again we have used Euclidean signature and
 the ellipses include curvature terms (derivatives of both metrics) which are ignored for the purpose of this discussion.
The inverse effective metric is given by
\ba
g^{\mu\nu}_{\rm eff}=\frac{{\lambda_g} \sqrt{g}g^{\mu \nu }+{\lambda_f} \sqrt{f}f^{\mu \nu}}{{\lambda_g} \sqrt{g}+{\lambda_f} \sqrt{f}}\,,
\ea
which, up to derivative corrections, leads to the standard cosmological constant contribution but this time associated to the effective metric
\ba
\mathcal{L}^{({\rm matter-loops})}_{1, \log }=M^4 \sqrt{g_{\rm eff}}\log(M/\mu) + \text{curvature corrections}\,.
\ea
Unfortunately this contribution $\sqrt{g_{\rm eff}}$ does not take the form of any of the allowed potential presented  in section \ref{sec:reviewMG}. We now expand this contribution to quadratic order in perturbations. For that we choose a ``vielbein-inspired" convention where the metric is given by
\ba
g\mn= \(\delta\mn + h\mn\)^2 \quad {\rm and} \quad
f\mn= \delta\mn  \,,
\ea
giving rise to
\ba
\mathcal{L}^{({\rm matter-loops})} \supset M^4 \sqrt{g_{\rm eff}}
= M^4 \Big(1+\frac {\lambda_g}{{\lambda_g}+{\lambda_f}} [h] + \frac {{\lambda_g}({\lambda_g}+2{\lambda_f})} {2({\lambda_g}+{\lambda_f})^2}\( [h]^2- [h^2]\) \\ - \frac{{\lambda_g} {\lambda_f}}{2({\lambda_g}+{\lambda_f})^2} [h^2] +\cdots\Big)\, \log(M/\mu) \,.\nn
\ea
While the first line includes terms which have the correct Fierz-Pauli structure to avoid the BD ghost, the second line brings a  new contribution   going as ${\lambda_g} {\lambda_f} [h^2]$ which is ghost-like. That contribution vanishes when either one of the couplings is absent (${\lambda_g}$ or ${\lambda_f}$ is dialed to zero) as one would expect.
Now considering the helicity-0 $\pi$ contribution $h\sim \p^2 \pi /\Lambda^3 $ and focusing in the decoupling limit $\mpl, M_{{\rm Pl}, f} \to \infty$, keeping the scale $\Lambda^3 = \mpl m^2 $ finite, we see the emergence of higher derivative terms
\ba
\mathcal{L}^{({\rm matter-loops})} \supset \frac{{\lambda_g} {\lambda_f}}{({\lambda_g}+{\lambda_f})^2} \frac{M^4}{\Lambda^6} \(\p^2 \pi\)^2+ \cdots\,,
\ea
leading to an Ostrogadski ghost \cite{Ostrogradsky} at an unacceptable low scale, namely
\ba
\label{eq:massGhost}
m^2_{\rm ghost}= \Lambda^6 / M^4\,.
\ea
In particular for a scalar field of mass $\Lambda$ we see a ghost at the scale $\Lambda$.
This means that the new potential contributions from the one-loop effective action already lead to a ghost within the decoupling limit itself.

\subsection{Classical ghost in the mini-superspace}
\label{sec:MSP1}

We can illustrate the presence of a ghost by looking at the special case of the mini-superspace.\footnote{We thank Andrew J.~Tolley for pointing this out.} Writing the respective metric in the mini-superspace as
\ba
\label{eq:gMSP}
\d s_g^2 = g\mn \d x^\mu \d x^\nu = -N^2(t) \d t^2+a^2(t)\d x^2\\
\d s_f^2 = f\mn \d x^\mu \d x^\nu = -\mathcal{N}^2(t) \d t^2+b^2(t)\d x^2\,,
\label{eq:fMSP}
\ea
and focusing on a time-dependent scalar field $\chi=\chi(t)$, the matter Lagrangian \eqref{eq:samefieldLag} simplifies to
\ba
\L_{\rm matter}=\frac12 \(\frac{{\lambda_g} a^3}{N}+\frac{{\lambda_f} b^3}{\mathcal{N}}\) \dot \chi^2 - \frac12 M^2 \({\lambda_g} a^3N+{\lambda_f} b^3 \mathcal{N}\) \chi^2\,.
\ea
The conjugate momentum associated to $\chi$ is given by
\ba
p_\chi= \(\frac{{\lambda_g} a^3}{N}+\frac{{\lambda_f} b^3}{\mathcal{N}}\) \dot \chi\,,
\ea
and the contribution to the Hamiltonian density from this matter field is then
\ba
\mathcal{H}_{\rm matter}= \frac 12 \frac{N \mathcal{N} }{{\lambda_g} a^3\mathcal{N}+{\lambda_f} b^3N} p_\chi^2 + \frac 12 M^2 \(\lambda_g a^3N+\lambda_f b^3 \mathcal{N}\)  \chi^2\,,
\ea
and is clearly not linear in neither lapses $N$ nor $\mathcal{N}$, unless either ${\lambda_g}$ or ${\lambda_f}$ vanishes. Since there is no shift in this mini-superspace example, the only way to generate a constraint capable of removing the BD ghost is if the lapses are linear. We can therefore immediately  conclude that the case of single field minimally coupling to both metrics simultaneously exhibits a BD ghost already at the classical level. If the scale associated with the ghost was large enough its effect would be unimportant for the low-energy effective field theory, however as we shall see below the ghost is already present in the decoupling limit and is thus an unacceptable way to couple to matter.

\subsection{Classical ghost in the decoupling limit}
\label{sec:DL1}

Another way to see the appearance of a ghost already at the classical level and to deduce its associated scale is to derive the decoupling limit. For that we first introduce the \stu fields to restore the broken copy of diffeomorphism invariance as mentioned in section~\ref{sec:review}. In bi-gravity, there is a choice associated with how to introduce these fields \cite{Fasiello:2013woa}. For massive gravity the \stu fields are there to promote the reference metric (say $f\mn$) to a tensor. Without loss of generality, we introduce the \stu by performing the following substitution in \eqref{eq:samefieldLag}
\ba
f\mn \to \tilde f\mn = f_{ab}\p_\mu \phi^a \p_\nu \phi^b\,.
\ea
We point out that this is a perfectly acceptable substitution even in the case where ${\lambda_g}=0$ and the field  only couples to the metric $f\mn$.

Next we split the \stu fields into their helicity-0 and -1 counterparts,
\ba
\phi^a=x^a-\frac{A^a}{m \mpl} -\frac{f^{ab}\p_b \pi}{\mpl m^2}\,,
\ea
where the fields $A$ and $\pi$ have been canonically normalised (their normalization is determined by their kinetic terms that enter in the potential interactions presented in section \ref{sec:reviewMG}).

We now take the decoupling limit where $\mpl, M_{{\rm Pl}, f} \to \infty$, while keeping the scale $\Lambda^3=\mpl m^2$ finite. In that limit it is sufficient to keep track of the helicity-0 $\pi$ contributions to the matter Lagrangian. When focusing on the contributions from \eqref{eq:samefieldLag} to the decoupling limit, it is thus safe to set $g\mn=\eta\mn$ and
\ba
\label{eq:fDL}
f\mn = \eta\mn \to \tilde f\mn = \(\eta\mn-\Pi\mn\)^2\,,
\ea
with $\Pi\mn \equiv \p_\mu \p_\nu \pi/\Lambda^3$, and we use the implicit notation that for any tensor $A\mn$, $A\mn^2\equiv A_{\mu\alpha}A^{\alpha}_{\ \nu}$, where in  the decoupling limit all indices are raised and lowered with respect to the Minkowski metric $\eta\mn$ unless specified otherwise. The resulting matter Lagrangian in the decoupling limit is thus
\ba
\label{eq:samefieldLagDec}
\L^{\rm dec}_{\rm matter} &=& {\lambda_g} \L^{\rm dec}_{g} +{\lambda_f} \L^{\rm dec}_{\tilde f}\\
&=&-\frac {\lambda_g} 2 \((\p \chi)^2 +M^2 \chi^2\)
-\frac {\lambda_f}2 \sqrt{- \tilde f} \( \tilde f^{\mu\nu}\p_\mu \chi \p_\nu \chi +M^2 \chi^2\)\,,
\ea
where $\tilde f^{\mu\nu}$ is the inverse of $\tilde f\mn$ and $\tilde f$ is the determinant of $\tilde f\mn$. Since $\tilde f\mn$ includes second derivatives of $\pi$, it is yet unclear whether the equations of motion with respect to both $\chi$ and $\pi$ are manifestly second order. However as mentioned previously, the theory has been shown to be ghost-free when either ${\lambda_g}$ or ${\lambda_f}$ vanishes and the field $\chi$ only couples to one metric. The case ${\lambda_f}=0$ is of course trivial since $\pi$ does not even enter. The case ${\lambda_g}=0$ is less so, and we shall study it first before dealing with the generic case.

\subsubsection{Coupling to only one metric: ${\lambda_g}=0$}
\label{sec:alpha0}

When ${\lambda_g}=0$, the Lagrangian \eqref{eq:samefieldLagDec} should propagate no BD (or Ostrogadski) ghost. To see this, we examine the equations of motion. The equation of motion for the matter field $\chi$ is
\ba
\mathcal{E}_\chi \equiv \Box_{\tilde f} \chi- M^2 \chi=0\,,
\ea
which is of course equivalent to
\ba
\label{eq:conserv}
\mathcal{E}_\chi^\nu \equiv  \nabla_{\mu}^{(\tilde f)} T^{\mu \nu}_{\tilde f}=0\,,
\ea
where $T^{\mu \nu}_{\tilde f}$ is the stress-energy tensor of $\chi$ associated with the metric $\tilde f\mn$
\ba
T^{\mu\nu}_{\tilde f}=\frac{2}{\sqrt{- \tilde f}} \frac{\delta \L^{\rm dec}_{\tilde f} }{\delta \tilde f\mn}
= \tilde f^{\mu\alpha}\tilde f ^{\nu \beta} \p_\alpha \chi \p_\beta \chi-\frac 12 \( \tilde f^{\alpha \beta} \p_\alpha \chi \p_\beta \chi+M^2 \chi^2\)\tilde f^{\mu\nu}\,.
\ea
More precisely, $\mathcal{E}_\chi^\mu\equiv \mathcal{E}_\chi \tilde f^{\mu\nu}\p_\nu \chi$, so $ \mathcal{E}_\chi =0$ trivially implies $\mathcal{E}_\chi^\mu=0$ which is a straightforward consequence of diffeomorphism invariance.

This equation for $\chi$ involves higher derivatives of $\pi$, but as an equation for $\chi$ it can be solved without any problem and without needing to require more than two initial conditions on the field $\chi$. So as long as the equation of motion for $\pi$ can be solved without requiring more than two initial conditions for $\pi$, the system will not suffer from any Ostrogadski or BD instability. As we shall see, this is the case when ${\lambda_g}=0$ (and trivially the case when ${\lambda_f}=0$) but no longer the case as soon as ${\lambda_g} {\lambda_f} \ne 0$.

The equation of motion with respect to $\pi$ is given by
\ba
\mathcal{E}_\pi= \mathcal{E}_{\pi} ^{\text{potential terms}}+ \mathcal{E}_\pi^{\rm matter}\,,
\ea
where $ \mathcal{E}_{\pi} ^{\text{potential terms}}$ is the contribution coming from the potential term and was shown in Ref.~\cite{deRham:2010ik} to be at most second order in derivative for the allowed potentials considered in section~\ref{sec:reviewMG}. $\mathcal{E}_\pi^{\rm matter}$ is defined as
\ba
\mathcal{E}_\pi^{\rm matter}= \frac{\delta  \L^{\rm dec}_{\rm matter} }{\delta \pi}= {\lambda_f}  \frac{\delta   \L^{\rm dec}_{\tilde f} }{\delta \pi}\,.
\ea
To be more precise, this contribution to the equations of motion for $\pi$ is
\ba
\label{eq:Epi}
\mathcal{E}_\pi^{\rm matter}&=&\frac{ {\lambda_f}}{\Lambda^3} \p_\mu \p_\nu \( \frac{\delta   \L^{\rm dec}_{\tilde f} }{\delta \tilde f_{\alpha \beta}} \frac{\delta \tilde f_{\alpha\beta}}{\delta \Pi\mn}\)\\
&=& - \frac{{\lambda_f}}{\Lambda^3}\p_\mu \p_\nu \(\sqrt{-\tilde f}\ T^{\mu \alpha  }_{\tilde f} \(\delta^{\nu}_{\alpha}- \Pi^{\nu}_{\alpha} \) \)\,.
\label{eq:Epi2}
\ea
Now from the equation of motion for $\chi$ \eqref{eq:conserv}, we infer that
\ba
\label{eq:partialT}
\nabla_{\mu}^{(\tilde f)} T^{\mu \alpha  }_{\tilde f}  =0 \quad \Rightarrow \quad \p_\mu \(\sqrt{-\tilde f}\ T^{\mu \alpha  }_{\tilde f}  \)= - \sqrt{-\tilde f} \ \tilde \Gamma^\alpha_{\mu \beta}  T^{\mu \beta  }_{\tilde f}\,,
\ea
where the $\tilde \Gamma$ are the Christoffel symbols associated with the metric $\tilde f\mn$.  This allows us to considerably simplify the contribution \eqref{eq:Epi2} to the equation of motion for $\pi$. After a few straightforward manipulations, we obtain,
\ba
\label{eq:Epis}
\mathcal{E}_\pi^{\rm matter}&=& {\lambda_f} \p_a \left[\sqrt{- \tilde f} \ T^{b c}_{\tilde f} R^a_{bc} \right] \,,
\ea
with the tensor $R^a_{bc}$ defined as
\ba
\label{eq:R}
R^a_{bc} \equiv \tilde \Gamma^\alpha_{bc}\(\delta^a_\alpha - \Pi^a_\alpha\)+ \p^a \Pi_{bc}\,.
\ea
This tensor only depends on $\pi$ and not on the matter field $\chi$. Upon closer examination, we can check that this tensor vanishes identically. First we notice that the Christoffel symbols are given by
\ba
\label{eq:Gamma}
\tilde \Gamma^\alpha_{bc}= - \tilde f^{\alpha \alpha'}\(\delta^\beta_{\alpha'}-\Pi^\beta_{\alpha'}\)\p _\beta \Pi_{bc}\,.
\ea
Second we recall that the metric $\tilde f\mn=\(\eta\mn-\Pi\mn\)^2$ so its inverse satisfies the relation
\ba
\label{eq:invMetric}
\(\delta^a_\alpha - \Pi^a_\alpha\) \tilde f^{\alpha \alpha'}\(\delta^\beta_{\alpha'}-\Pi^\beta_{\alpha'}\)\equiv \eta^{a\beta}\,.
\ea
Plugging these two relations \eqref{eq:Gamma} and \eqref{eq:invMetric} into the expression \eqref{eq:R} for $R$, we find
\ba
R^a_{bc}\equiv -  \p^a \Pi_{bc} + \p^a \Pi_{bc}\equiv 0\,.
\ea
The contribution from the matter Lagrangian to the equation of motion for $\pi$ therefore vanishes identically, $\mathcal{E}_\pi^{\rm matter}\equiv 0$ and the equation of motion for $\pi$ is thus the standard one without the contribution from the matter field, which is manifestly second order. We therefore  do not need to specify more than two initial conditions to solve the equation of motion for $\pi$ and the system with ${\lambda_g}=0$ is manifestly free from the BD ghost (in the decoupling limit). This is consistent with the results of Ref.~\cite{Hassan:2011hr} where it was shown that if a matter field couples minimally to a single metric the theory was free of the BD ghost in all generality.

\subsubsection{Minimal coupling to both metrics: ${\lambda_g} \ne 0$}

Next we consider the case where the matter field $\chi$ minimally couples to both metrics as in \eqref{eq:samefieldLagDec}. In that case, the equation of motion is still given by \eqref{eq:Epi} or \eqref{eq:Epi2}, but the equation of motion for $\chi$ is modified and the stress-energy tensor $T^{\mu\nu}_{\tilde f}$ is no longer conserved.

The equation of motion for $\chi$ now reads
\ba
\mathcal{E}_\chi \equiv {\lambda_g} \(\Box-M^2\chi\)+{\lambda_f} \sqrt{-\tilde f}\,\(\Box_{\tilde f} \chi- M^2 \chi\)=0\,,
\ea
where $\Box$ is the flat space-time d'Alembertian, $\Box=\eta^{\mu\nu}\p_\mu \p_\nu$. This equation is no longer equivalent to the conservation of the individual stress-energy tensors, but only to the sum of them,
\ba
\mathcal{E}_\chi^\mu \equiv {\lambda_g} \p_\mu T^{\mu\nu}_{\eta} +{\lambda_f} \nabla_{\mu}^{(\tilde f)} T^{\mu \nu}_{\tilde f} =0\,.
\ea
This means that the rest of the previous argument when ${\lambda_g}$ vanished goes through, with the only difference that \eqref{eq:partialT} should be replaced by
\ba
{\lambda_f} \p_\mu \(\sqrt{-\tilde f}\ T^{\mu \alpha  }_{\tilde f}  \)= - {\lambda_f} \sqrt{-\tilde f} \ \tilde \Gamma^\alpha_{\mu \beta}  T^{\mu \beta  }_{\tilde f} - {\lambda_g} \sqrt{-\tilde f} \  \p_\mu T^{\mu\alpha}_{\eta}\,.
\ea
This leads to a new contribution in the equation for $\pi$,
\ba
\label{eq:Epi2}
\mathcal{E}_\pi^{\rm matter}&=& \frac{{\lambda_g}}{\Lambda^3} \(\p_\nu \(\sqrt{-\tilde f} \p_\mu T^{\mu \alpha}_{\eta}\)\) \(\delta^\nu_\alpha -\Pi^\nu_\alpha\)-\frac{{\lambda_g}}{\Lambda^3} \p_\mu T^{\mu\alpha}_\eta \p_\alpha \Box \pi\,,
\ea
which does not vanish unless either ${\lambda_g}=0$ or ${\lambda_f}=0$ (in which case $ \p_\mu T^{\mu \alpha}_{\eta}=0$). As a result, the equation of motion for $\pi$ contains higher derivative terms which cannot be removed by invoking the equation of motion with respect to $\chi$ unless ${\lambda_g} {\lambda_f}=0$. To conclude we have shown via three different methods (two classical arguments and one quantum) that a single field cannot minimally couple to both metrics $g$ and $f$ simultaneously without leading to a ghost at an unacceptable low scale.

We now build on the lessons learnt here and propose a different way to couple a single matter field to both metrics simultaneously  via a single effective metric.

\section{Effective metric}
\label{sec:effmet}

\subsection{One--loop requirement}

In principle, one should start by looking for couplings to matter which are free of the BD ghost at the classical level. However, to connect with our previous derivations, we start here instead by asking a slightly different question: For what kind of coupling would the one-loop corrections from virtual matter fields take the form of the allowed potential ?
As we shall see, this is a much simpler and straightforward requirement to satisfy. This will allow us to find an effective metric and further argue that it is free of the BD ghost (at least in some limits).

Let us consider a single matter field $\chi$ (taken to be a scalar field for concreteness, but the exact nature of that field is irrelevant to this argument) minimally coupled to an effective metric $\gd$ as follows,
\ba
\label{eq:Lmatter geff}
\L_{\rm matter}= -\frac12 \sqrt{-\gd}\(\gd^{\mu\nu}\p_\mu\chi \p_\nu \chi+M^2 \chi^2\)\,,
\ea
where the effective metric $\gu\mn$ is built out of $g\mn$ and $f\mn$. As shown previously, the one-loop effective action from matter loops will lead to a contribution going as
\ba
\L_{\rm 1-loop} = M^4 \sqrt{-\gd} +\text{curvature corrections}\,.
\ea
If $\gd$ is built out of both metrics $g\mn$ and $f\mn$, the one-loop contribution $\sqrt{-\gd}$ will typically lead to a ghost at an unacceptable low scale, {\it unless} $\sqrt{-\gd}$ happens to take the form of one of the allowed potentials introduced in section~\ref{sec:reviewMG}. Such an acceptable class of effective metrics is for instance given by
\ba
\label{eq:geff}
\gu\mn=\alpha^2 g\mn +2\alpha \beta\ g_{\mu\alpha} X^\alpha_{\ \nu} +\beta^2 f\mn\,,
\ea
where we recall that $X=\sqrt{g^{-1}f}$ and $\alpha$ and $\beta$ are two arbitrary real dimensionless parameters.

While the quantity $X^\alpha_{\ \nu}$ is not symmetric, we can show that the tensor $Y_{\mu\alpha}X^\alpha_{\ \nu}$ that enters the effective metric is symmetric. Working in matrix notation with $X\equiv \sqrt{g^{-1}f}$ and $Y\equiv g X$, we have
\ba
Y=g X = g X g^{-1} g\,.
\ea
$gX g^{-1}$ is nothing other than a similarity transformation on the matrix X, and by properties of the similarity transformations, it follows that $g X g^{-1} g = \sqrt{g g^{-1}f g^{-1}}=\sqrt{f g^{-1}}=X^T$ (using the fact that both $g$ and $f$ are symmetric). This implies that $Y$ itself is symmetric,
\ba
Y=X^T g= X^T g^T=Y^T\,.
\ea

Clearly when $\alpha=0$ or $\beta=0$, the field only couples to one metric and $\sqrt{-\gd}$ is simply the cosmological constant associated to that metric. More generally one can show that 
\ba
\label{eq:detgeff}
\sqrt{-\gd}= \sqrt{-\det g}\  \det\(\alpha +\beta X\)\,,
\ea
which is precisely of the form of the acceptable classes of potentials as shown in \eqref{eq:det1}. When $\alpha=0$ or $\beta=0$, we see that the contribution is nothing else but the standard cosmological constant for either the metric $g\mn$ or the metric $f\mn$, and the mass term (or interactions between the metrics $g\mn$ and $f\mn$) would not get destabilised by integration of matter loops. The mass parameter therefore remains technically natural as found in \cite{deRham:2012ew,deRham:2013qqa}.
On the other hand when $\alpha \beta \ne 0$, quantum corrections from matter fields do destabilise the scale $m$ which is no longer technically natural, unless only massless fields couple to that effective metric.

\subsection{Coupling to other fields}
The previous argument is not related to the nature of the matter field in \eqref{eq:Lmatter geff}. Had one considered a massive or massless vector field the result would have remained the same. Similarly, when coupling to spinors, one should work in the vielbein formalism with $\gu\mn = e_{\rm eff}{}^a_\mu e_{\rm eff}{}^b_\nu \eta_{ab}$, and the spinors coupled to the effective vielbein $e_{\rm eff}$ constructed out of the vielbein from both the $g$ and the $f$ metric. Irrespectively, the contribution to the metric potential at one--loop from integrating out any of these matter fields is proportional to $\sqrt{-\gd}$.

Similarly one could consider more complicated couplings to matter of the form
\ba
\L_{\rm matter}= \sqrt{-\gd}L\(\gu\mn, \chi, \p_\mu \chi, \nabla_\mu^{\rm (eff)} \nabla_\nu^{\rm (eff)} \chi, \cdots, A_\mu, \cdots\)\,,
\label{eq:Lmatter gen eff}
\ea
as long as these couplings are covariant with respect to the metric $\gd$, the one--loop effective action contribution from integrating out these matter fields is also covariant with respect to $\gd$ and cannot be anything else but $\sqrt{-\gd}$.

Finally this result also generalises to any number of loops following the same argument as in Ref.~\cite{deRham:2013qqa} provided only matter fields run in the loops.\\

In what follows we show that the coupling \eqref{eq:Lmatter geff} does not lead to any BD ghost in the mini-superspace approximation and in the decoupling limit. We focus on the Lagrangian \eqref{eq:Lmatter geff} for concreteness but all the results presented below go through trivially in the same way when considering a more general coupling \eqref{eq:Lmatter gen eff}, provided we only ask the question about the BD ghost (depending on the number of derivatives involved in \eqref{eq:Lmatter gen eff}, that matter Lagrangian may include ghosts of its own, even if we replaced $\gd$ by $g\mn$ and considered the GR case. This is of course a separate question independent of the coupling to $\gd$.)

\subsection{Absence of ghost in the mini-superspace}
\label{sec:MSS}
First we consider the same mini-superspace configuration as we did in section~\ref{sec:MSP1}. This is already a non-trivial check since section~\ref{sec:MSP1} was already sufficient to diagnose a ghost for the Lagrangian~\eqref{eq:samefieldLag}. Of course the mini-superspace approximation corresponds to a generic FLRW one and thus applies to any FLRW solutions one may find in the theory, as will be the case in section \ref{sec:FRW}.

Keeping the same form for the metrics $g\mn$ and $f\mn$ as in \eqref{eq:gMSP} and \eqref{eq:fMSP},  the effective metric is simply given by
\ba
\label{eq:geffMSP}
\d s_{\gd}^2 = \gu\mn \d x^\mu \d x^\nu = -\(\alpha N(t)+\beta \mathcal N(t)\)^2 \d t^2+\(\alpha a(t)+\beta b(t)\)^2\d x^2\,.
\ea
Focusing again on a time-dependent scalar field $\chi=\chi(t)$, the matter Lagrangian \eqref{eq:Lmatter geff} now simplifies instead to
\ba
\L_{\rm matter}=\frac12 \(\alpha a +\beta b\)^3 \(\frac{1}{\alpha N+\beta \mathcal N} \dot \chi^2 -  M^2 \(\alpha N+\beta \mathcal N\) \chi^2\)\,.
\ea
The conjugate momentum associated to $\chi$ is thus given by
\ba
p_\chi=  \frac{\(\alpha a +\beta b\)^3}{\alpha N+\beta \mathcal N} \dot \chi\,,
\ea
and the contribution to the Hamiltonian density from this matter field is then
\ba
\mathcal{H}_{\rm matter}= \frac 12 \frac{\alpha N+\beta \mathcal N }{\(\alpha a +\beta b\)^3} p_\chi^2 + \frac 12 M^2 \(\alpha a +\beta b\)^3\(\alpha N+\beta \mathcal N\)   \chi^2\,.
\ea
The mini-superspace Hamiltonian is thus clearly linear in both lapses $N$ and $\mathcal{N}$, independently of the values of $\alpha$ and $\beta$. Once again, since there is no shift in this mini-superspace example, the linearity in the lapse is sufficient to generate a  constraint capable of removing the BD ghost. So at least in the mini-superspace, the matter Lagrangian \eqref{eq:Lmatter geff} does not excite any BD ghost, even though it couples the field $\chi$ to both metrics $g\mn$ and $f\mn$ simultaneously.

\subsection{Absence of ghost in decoupling limit}
\label{subsec:ghostdl}
Next, we turn to the proof of the absence of ghost in the decoupling limit. By itself this proof is not sufficient to ensure that there is no ghost at all (even classically). However, an absence of ghost in the decoupling limit means that the scale associated with the ghost is larger than $\Lambda$, and therefore it does make sense to consider that theory with a cut-off which is at least $\Lambda$, if not higher if there is a Vainshtein mechanism at work by which non-linearities become important.

To derive the decoupling limit associated with \eqref{eq:Lmatter geff} we proceed as in section~\ref{sec:DL1} and set  $g\mn=\eta\mn$ and
\ba
f\mn = \eta\mn \to \tilde f\mn = \(\eta\mn-\Pi\mn\)^2\,,
\ea
with again $\Pi\mn \equiv \p_\mu \p_\nu \pi/\Lambda^3$. In this limit, the effective metric is then given by
\ba
\gu\mn= \left[(\alpha+\beta)\eta\mn-\beta \Pi\mn\right]^2\,.
\ea
With this effective metric in the decoupling limit, we can now proceed precisely as in section~\ref{sec:alpha0}.
The equation of motion for the matter field $\chi$ is given by
\ba
\label{eq:conserv2}
\mathcal{E}_\chi^\mu \equiv  \nabla_{\mu}^{(\gd)} T^{\mu \nu}_{\gd}=0\,,
\ea
where $T^{\mu \nu}_{\gd}$ is the stress-energy tensor of $\chi$.
Here again, this equation for $\chi$ involves higher derivatives of $\pi$, but is a well-defined equation for $\chi$ that can be solved by setting only two initial conditions for the field $\chi$.

Next, the contribution from \eqref{eq:Lmatter geff} to the equation of motion with respect to $\pi$ is given by
\ba
\label{eq:Epieff}
\mathcal{E}_\pi^{\rm matter}= - \frac{1}{\Lambda^3}\p_\mu \p_\nu \(\sqrt{-\gd}\ T^{\mu \alpha  }_{\gd} \( (\alpha+\beta) \delta^{\nu}_{\alpha}- \beta \Pi^{\nu}_{\alpha} \) \)\,.
\ea
Now proceeding precisely as in section~\ref{sec:alpha0} and using the relation  \eqref{eq:conserv2} in the previous expression for $\mathcal{E}_\pi^{\rm matter}$, we find similarly as before,
\ba
\label{eq:Epieff2}
\mathcal{E}_\pi^{\rm matter}&=& -  \p_a \left[\sqrt{- \gd} \ T^{b c}_{\gd} R^a_{bc} \right] \,,
\ea
where the tensor $R^a_{bc}$ is now defined as
\ba
\label{eq:Reff}
R^a_{bc} \equiv  \Gamma^\alpha_{bc}\( (\alpha+\beta)\delta^a_\alpha - \beta\Pi^a_\alpha\)+ \beta \p^a \Pi_{bc}\,.
\ea
Where the $\Gamma^\alpha_{bc}$ are the Christoffel symbols associated with the effective metric $\gd$. They satisfy the relation
\ba
 \Gamma^\alpha_{bc}\( (\alpha+\beta)\delta^a_\alpha - \beta\Pi^a_\alpha\) = -\beta \p^a \Pi_{bc} \,,
\ea
which ensures the following identity
\ba
R^a_{bc}\equiv 0\,.
\ea
So just like in the case where the field $\chi$ only coupled to one metric, we see here that the contribution from the matter Lagrangian to the equation of motion for $\pi$ vanishes identically, $\mathcal{E}_\pi^{\rm matter}\equiv 0$ and the equation of motion for $\pi$ is thus the standard one without the contribution from the matter field, which is manifestly second order.
As a result, there  is no ghost in the decoupling limit when the field $\chi$ couples to both metrics simultaneously via the effective metric $\gd$.

This is an important result by itself. Even if a careful constraint analysis ended up signaling the existence of a ghost, the absence of BD ghost in the decoupling limit means that the scale associated with the ghost is higher than $\Lambda$ and therefore the theory is consistent at that scale. The existence of a ghost in the full theory would inevitably set a cut-off scale $\Lambda_c$ larger than $\Lambda$.

\subsection{Re-emergence of the ghost}
\label{sec:remergenceGhost}

Performing a more careful ADM analysis in all generality (\ie\ beyond the mini-superspace approximation performed previously), we find,
after integrating out the shift in the Hamiltonian,  an operator of the form
\ba
\mathcal{H}\supset \frac{1}{m^2 \mpl^2}\frac{\alpha^2 \beta^2}{(\alpha+\beta)^2} (\p_i \chi)^2 p_\chi^2 N^2\,,
\ea
which is quadratic in the lapse and destroys the constraint that would otherwise project out the BD ghost, (see Ref.~\cite{deRham:2014fha} for further details.)
The mass of the associated ghost depends on the behaviour of the scalar field, and the background configuration but the overall associated scale is given by $\Lambda_2^2=\mpl m$.
Inspired by the \stu formalism, we may set $N\sim \ddot \pi/(\mpl m^2)$. Assuming that this was the most relevant operator in determining the scale of the ghost, and assuming that this substitution is appropriate at these scales this would lead to a mass of the ghost going as $m_{\rm ghost}\sim m^3 \mpl^2/(\p_i\chi) p_\chi$. For homogeneous backgrounds, $\p_i \chi=0$ and that scale goes to infinity which is consistent with the result found on FLRW in section \ref{sec:MSS} where no ghost is present. Whether the operator identified here is the most important one for the discussion and whether the substitution $N\sim \ddot \pi/(\mpl m^2)$ is appropriate is unclear at this stage (see Ref.~\cite{deRham:2011rn} for a discussion of this point) and a more detailed analysis should be performed. At this level, the analysis performed so far strongly suggests the existence of a ghost for highly anisotropic solutions.

\section{Consequences for Cosmology---FLRW solutions}
\label{sec:FRW}

\subsection{Exact FLRW solutions}

The no-FLRW solution for massive gravity result derived in Ref.~\cite{DAmico:2011jj} relied on the fact that the \stu equation of motion imposes a too strong constraint on the scale factor. When at least one physical matter field couples to the effective metric \eqref{eq:geff} the \stu enter the matter Lagrangian and the previous results are thus modified, allowing for the possibility of exact FLRW solutions in massive gravity (with flat reference metric and no spatial curvature). These couplings would also have interesting consequences for bi-gravity, but the cosmology of these couplings is beyond the scope of this work and in what follows we merely point out the existence of consistent FLRW solutions in massive gravity.

Starting with the homogeneous and isotropic ansatz for the metric and fixing unitary gauge for the \stu fields\footnote{One could also fix a gauge for the metric and keep the \stu fields in, which is of course completely equivalent procedure  \cite{DAmico:2011jj}.},
\ba
\d s_g^2=- N(t)^2\d t^2 +a(t)^2 \d\vec{x}{}^2\,, \quad \quad{\rm and}\quad \quad \phi^a=x^\mu \delta^a_\mu \,.
\ea
The effective metric then takes the form
\ba
\d s_{\gd}^2=-\tilde N(t)^2  \d t^2 + \tilde a(t)^2  \d\vec{x}{}^2\,, \quad {\rm with}\quad
\tilde N(t)=\alpha N(t)+\beta\quad{\rm and}\quad \tilde a(t) = \alpha a(t) + \beta \,,
\ea
which corresponds to $\mathcal{N}=1$ and $b=1$ in Eq.~\eqref{eq:geffMSP}. In what follows we consider  the case of a single scalar field $\chi$ with potential $V(\chi)$ minimally coupled to the effective metric $\gu\mn$ as well as a standard perfect fluid with energy density $\rho$ and pressure $p$ coupled covariantly to the dynamical metric $g\mn$. We focus on the case of a canonical scalar field  $\chi$ for simplicity but the results of this section are valid for much more general types of matter (so long as it is described by a field that does include a kinetic term). We refer to appendix~\ref{app:appendix} for a more general analysis.

As will become clear from our resulting Friedmann equation, the fields that couple to the effective metric end up contributing in a very peculiar way to the evolution of the Universe. For this reason we break the equivalence principle and do not couple all the fields to the effective metric. For the fields that couple to the standard metric $g\mn$, a standard perfect fluid approach is a sufficient and acceptable description of their physics. However for the fields that couple to the effective metric, it is important to properly understand how the lapse really enters (or how the \stu fields enter if one was not working in unitary gauge -- this is at the very core of the no-FLRW solution result for massive gravity.) For this reason we keep a field description of the matter fields that couple to the effective metric. One could be tempted to describe the physical fields that couple to the effective metric using a  perfect fluid approximation with an energy density and a pressure which depend neither on the lapse nor on the \stu fields (see for instance Ref.~\cite{Solomon:2014iwa}). However this  would be an incorrect description of the system as it would not properly account for how the lapse or the \stu fields enter their Lagrangian and interact with these fields.

As a result we use the following Lagrangian as a representative toy--model
\ba
\label{eq:totalmatter}
\mathcal{L}=\L_{\rm mGR}+\mathcal{L}_{\rm  \rho}(g,\rho)+\mathcal{L}_{\rm \chi}(g_{\rm eff},\chi)\,,
\ea
with $\mathcal{L}_{\rm  \rho}(g,\rho) = -\sqrt{-g} \rho(a)$ satisfying the conservation of energy
\ba
a \rho'(a)=-3\(\rho+p\)\,,
\ea
and
\ba
\label{eq:chiL}
\mathcal{L}_{\rm \chi}(g_{\rm eff},\chi)=\sqrt{-\gd}\(-\frac 12 \gd^{\mu\nu} \p_\mu \chi \p_\nu \chi-V(\chi)\)\,.
\ea
Using the previous homogeneous and isotropic ansatz with $\chi=\chi_0(t)$ we obtain
\ba
\label{Ltotal}
\L=3 \mpl^2 \left[- \frac{a \dot a^2}{N}-\frac{m^2}{2}a(1-a)\(a+(1-2a)N\)
\right] - N a^3 \rho(a)
+\tilde a^3 \tilde N \left[\frac{\dot \chi_0^2}{2 \tilde N^2}- V(\chi_0)\right]\,,
\ea
where for simplicity we focus on the massive gravity case where $\alpha_n =\delta_{n,2}$ in Eq.~\eqref{eq:LagMGR}.
Notice that since we have chosen $\alpha_{0,1}=0$ this Lagrangian contains no cosmological constant\footnote{In massive gravity with Minkowski reference metric, the notion of cosmological constant is simply determined and setting $\alpha_{0,1}=0$ ensures its absence.}.
The first two terms  in \eqref{Ltotal} (\ie\ the term in square bracket and the $\rho$ term)  correspond to the standard massive gravity action, while the last term in square brackets includes the coupling to the effective metric $\gd$.

Varying this action with respect to the lapse, we obtain the would-be Friedmann constraint in GR,
\ba
\label{eq:Fried1}
 3\mpl^2 H^2=\rho+\frac{3m^2\mpl^2}{2a^2}(1-a)(1-2a)+\alpha\(\frac{\tilde a}{a}\)^3 \left[\frac{\dot \chi_0^2}{2 \tilde N ^2}+ V(\chi_0)\right]\,,
\ea
where $H$ is the standard Hubble parameter $H=\dot a/ a N$.
The consistency of this constraint with the evolution equation imposed by varying with respect to the scale factor and the $\chi$ equation of motion imposes
\ba
\label{eq:ConsMG}
m^2 \mpl^2 (a N)^{-1}\p_t \(a^2-a^3\) = 2\alpha \beta  \tilde a^2 H \(\frac{\dot \chi_0^2}{2 \tilde N^2}- V(\chi_0) \)\,.
\ea
In the case where all the matter fields uniquely couples to the standard metric rather than to the effective metric (\ie\ when $\beta=0$), or when the matter couples directly to the reference metric (\ie\ when $\alpha=0$), the right hand side vanishes and we immediately recognise the constraint which prevents the existence of exact FLRW solutions in massive gravity. This constraint is now alleviated, and can instead be read as an equation for the lapse $N$. Solving Eq.~\eqref{eq:ConsMG} for $N$ and substituting in
Eq.~\eqref{eq:Fried1}, we obtain the modified Friedmann equation
\ba
3\mpl^2 H^2=\rho+\frac{m^2\mpl^2 }{2\beta a^2}\(5\beta+2(\alpha-6 \beta)a -3(\alpha-2 \beta)a^2\)+2\alpha \(\frac{\tilde a }{a}\)^3V(\chi_0)\,.
\ea
Interestingly, the Hubble parameter is no longer (directly) affected by the kinetic term of the field $\chi$ coupling to $\gd$, but it is determined by the energy density of the fields living on the dynamical metric. It is quite possible that the presence of terms scaling as $a^{-1}$  or $a^{-2}$ might be ruled out by observations, although we bare in mind that these terms not only decay but are also suppressed by the scale $m^2$. Moreover, we have only considered a specific example here $\alpha_n=\delta_{n,2}$ as a proof of principle that exact FLRW solutions could exist. We emphasize that by setting $\alpha_{0,1}=0$ we have ensured the absence of cosmological constant and we see that we can still have $H\ne0$ even when $\rho=0$ and in the absence of a cosmological constant. The reason for that is the well-know fact that the graviton mass terms can act as a source for the acceleration of the Universe, corresponding to self-accelerating solutions.
This should be investigated with care and a full analysis is beyond the scope of this work.

\subsection{Linear Hamiltonian analysis about the exact FLRW solution}

As shown in section \ref{sec:remergenceGhost}, while this coupling to matter does not lead to any ghost in the decoupling limit, the BD ghost still appears below the Planck scale although above the strong coupling scale $\Lambda$. In order to trust the exact FLRW solutions found previously, it is therefore essential to establish whether or not this background excites the BD ghost.

We have already established the absence of BD ghost in the mini-superspace approximation. This means that for the FLRW configurations considered in the previous section, there are {\bf no} BD-ghost type of operators. The solution we have found would therefore be exactly the same as the one we would have obtained had we surgically removed all the ghostly-type of operators in the full theory. Stated differently, the existence of exact FLRW solutions is not relying on the existence of ghost in the full theory (since the BD ghost is absent in the FLRW case), but rather on the existence of new healthy-type of operators.

To push the analysis even further, we perform a  Hamiltonian ADM analysis for perturbations about  generic FLRW backgrounds (independently of whether or not they are classical solutions). The reference metric remains flat, $f\mn=\eta\mn$, while the dynamical metric takes the form
\ba
\label{pertgFRW}
g^{00}= -(N(t)+\delta N(t, x^i))^{-2}\,, \quad g_{0i}= N_i(t, x^i) \quad {\rm and}\quad g_{ij}= a(t)^2 \delta_{ij}+\gamma_{ij}(t, x^i) \,,
\ea
and the scalar field coupled to the effective metric
\ba
\chi(t,x^i)=\chi_0(t)+\delta \chi (t, x^i)\,,
\ea
where $\delta N$, $N_i$, $\gamma_{ij}$ and $\delta \chi$ are all considered to be first order in perturbations.

When performing the Hamiltonian constraint analysis, the only potentially problematic piece is that arising from the matter Lagrangian coupling the effective metric with the field $\chi$ derived from \eqref{eq:chiL}. Its contribution to the Hamiltonian density is
\ba
\mathcal{H}_\chi(\gd, \chi)=\sqrt{-\gd} \( \frac 12(-\gd^{00})^{-1} \(\frac{p_\chi}{\sqrt{-\gd}} + \gd^{0i}\p_i \chi\)^2
+\frac 12   \gd^{ij} \p_i \chi \p_j \chi + V(\chi)\)\,,
\ea
where $p_\chi$ is the conjugate momentum associated with $\chi$, $p_\chi=\p \L_\chi / \p \dot \chi$.

First notice that the effective metric has been built so that $\sqrt{-\gd}$ always keeps the correct structure so there can be no problem from that term. Indeed, with the ansatz \eqref{pertgFRW}, the determinant of the effective metric takes the form
\ba
\sqrt{-\gd}=\tilde a^3 \tilde N+\alpha \tilde a^2 \(\tilde a \delta N +\frac a 2 \tilde N [\gamma]\)+\frac {\alpha^2} 2 a \tilde a^2 \delta N [\gamma]
+\frac{\alpha^2}{8}a^2 \tilde a \tilde N \([\gamma]^2-[\gamma^2]\) \\
-\frac \alpha 8 \tilde a^2 a \tilde N [\gamma^2]-\alpha \beta \frac{\tilde a^2}{2(a+N)}N_i^2+\text{higher order perturbations}\,,\nn
\ea
where $[\gamma]=a^{-2}\delta^{ij}\gamma_{ij}$, $[\gamma^2]=a^{-4}\delta^{ij}\delta^{k\ell}\gamma_{ik}\gamma_{j\ell}$ and $N_i^2=a^{-2}\delta^{ij} N_i N_j$.

As expected, $\sqrt{-\gd}$ is linear in the perturbed lapse $\delta N$ and there can be no mixing between the lapse and the shift. So up to second order in perturbations the contributions from the two last terms in the Hamiltonian, namely from $\sqrt{-\gd} \gd^{ij} \p_i \chi \p_j \chi $ and from $\sqrt{-\gd} V(\chi) $ are manifestly linear in the perturbed lapse $\delta N$.

Next we turn to the only potentially problematic term in the Hamiltonian, namely the term $\sqrt{-\gd} (-\gd^{00})^{-1} \(\frac{p_\chi}{\sqrt{-\gd}} + \gd^{0i}\p_i \chi\)^2$ and check that up to second order in perturbations it remains linear in the perturbed lapse. Indeed one can easily check that $(g^{00}\sqrt{-\gd})^{-1}$ is linear in the perturbed lapse and cannot involve a mixing between the shift and the lapse. Since  $\p_i \chi$ are already first in perturbation, there can be no issue from these terms. As a result the perturbed Hamiltonian remains linear in the lapse.

To that order, the perturbed lapse thus generates a constraint which is  sufficient to remove half a physical degree of freedom (half the BD ghost). Since there cannot be half a physical degree of freedom in this configuration, there must be a secondary second class constraint which removes the other half of the BD ghost. We conclude that there is no BD ghost on the FLRW configuration presented here.

\section{Other consistent couplings to matter}
\label{sec:OtherCoupling}

As seen in the mini-superspace and decoupling limit analyses performed in sections \ref{sec:MSP1} and \ref{sec:DL1}, one of the issues when a scalar field couples covariantly to both metrics directly is that its kinetic term re-introduces the BD ghost.  We may thus consider a slightly different situation where the kinetic term of the matter field is dictated by only one metric (be it $g\mn$  or $f\mn$) while the potential term can see a different metric. As a specific yet educative example, we may consider the following matter coupling,
\ba
\label{eq:Othercoupling}
\L_{\rm matter}=-\frac 12 \sqrt{-g}g^{\mu\nu}\p_\mu\chi \p_\nu \chi - \sqrt{-f}\ V(\chi)\,.
\ea
In what follows we show that such a coupling is free from the BD ghost classically (at least in some limits) but that the ghost reappears at the quantum level at an unacceptable low-scale.

\subsection{Mini-superspace}
Using the same ansatz as in Eqs.~\eqref{eq:gMSP} and \eqref{eq:fMSP} with a time-dependent scalar field $\chi=\chi(t)$, the contribution to the Hamiltonian from the matter Lagrangian \eqref{eq:Othercoupling} is given by
\ba
\mathcal{H}_{\rm matter}= \frac{N}{2 a^3}p_\chi^2+b^3 \mathcal{N}V(\chi)\,,
\ea
where $p_\chi$ is the conjugate momentum  associated with $\chi$, $p_\chi=N a^{-3} \dot \chi$.  We immediately see that the Hamiltonian is linear in both lapses $N$ and $\mathcal{N}$ which ensures the propagation of the appropriate constraints.

\subsection{Decoupling limit}

Now turning to the decoupling limit, we can perform the same analysis as in section \eqref{sec:DL1}, and set $g\mn=\eta\mn$ and replace $f\mn$ by $(\eta\mn-\Pi\mn)^2$ as in Eq.~\eqref{eq:fDL}. In the decoupling limit, the contribution from the matter Lagrangian \eqref{eq:Othercoupling} is therefore
\ba
\L^{\rm dec}_{\rm matter}= -\frac 12 (\p \chi)^2- \det \(\eta\mn-\Pi\mn\) V(\chi)\,,
\ea
where indices are raised and lowered with respect to flat Minkowski metric. The couplings between the two fields $\chi$ and $\pi$ are nothing else but a specific case of a generalised bi-Galileon \cite{Padilla:2010de,Deffayet:2011gz} and are thus free of Ostrogadski instability \cite{Ostrogradsky,Chen:2012au}. The decoupling limit is thus also free from the BD ghost at the classical level. 
We now turn to the quantum corrections to the graviton potential mediated by such a coupling.

\subsection{Quantum corrections}

To compute the quantum corrections to the graviton potential generated by virtual loops of the matter field $\chi$, we focus on the case of a pure mass term for the scalar field,  $V(\chi)=\frac 12 M^2 \chi^2$. In that case the running contributions to the one--loop effective action are given by
\ba
\L_{1,  \log}^{\rm (matter\ loops)} = M^4 \frac{\det f}{ \sqrt{\det g}} \log (M/\mu)+\text{curvature corrections}\,,
\ea
where we only focus on the graviton potential. This new contribution to the potential does not take the form of one of the ghost-free contributions given in \eqref{sec:reviewMG}. As a result, quantum corrections will induce a BD ghost even if it is not present at the classical level. The emergence of terms which do not respect the ghost-free structure of the potential at the quantum level should not come as a surprise, after all quantum corrections from the graviton itself already lead to such a detuning \cite{deRham:2013qqa}. However, as seen in Ref.~\cite{deRham:2013qqa} the quantum corrections from the graviton itself enter at a scale which makes them unimportant for what concerns the theory. In the case presented here, however, the amplitude of the quantum corrections is mediated by the scale $M$ and the mass of the ghost it generates is given by
\ba
m^2_{\rm ghost}= \frac{\Lambda^6}{M^4}\,,
\ea
similarly as in \eqref{eq:massGhost},  where $\Lambda$ is the strong coupling scale $\Lambda=(m^2 \mpl)^{1/3}$. Here again, for a scalar field of mass $\Lambda$ we see a ghost already at the scale $\Lambda$.

\subsection{FLRW solutions}

For completeness, we now turn to the existence of exact FLRW solutions in massive gravity when the coupling \eqref{eq:Othercoupling} is considered. We follow the same prescription as in section~\ref{sec:FRW}.

The Lagrangian is given by
\ba
\L=3 \mpl^2 \left[- \frac{a \dot a^2}{N}+\frac{m^2}{2}a(1-a)\(a+(1-2a)N\)
\right] - N a^3 \rho(a)+\frac{a^3}{2N}\dot \chi_0^2 - V(\chi_0)\,,
\ea
where again for simplicity we focus on the massive gravity case where $\alpha_n = \delta_{n,2}$ in Eq. \eqref{eq:LagMGR}.

Varying this action with respect to the lapse, we obtain the would-be Friedmann constraint in GR given now by
\ba
\label{eq:Fried2}
 3\mpl^2 H^2=\rho-\frac{3m^2\mpl^2}{2a^2}(1-a)(1-2a)+\frac {1}{2N^2}\dot \chi_0^2\,.
\ea
The consistency of this constraint with the evolution equation imposed by varying with respect to the scale factor and the $\chi$ equation of motion imposes now
\ba
\label{eq:ConsMG3}
3 m^2 \mpl^2 a^2 N H (2-3 a)=2 V'(\chi_0) \dot \chi_0\,.
\ea
As expected it is essential to have a potential for $\chi$ in order to obtain a non-trivial equation.
Substituting Eq.~\eqref{eq:ConsMG3} into Eq.~\eqref{eq:Fried2}, we obtain the very peculiar yet non trivial Friedmann equation
\ba
3\mpl^2 H^2=\rho-\frac 32 \frac{m^2 \mpl^2}{a^2}(1-a)(1-2a)+\frac{9 m^4 \mpl^4}{8 V'(\chi_0)^2}a^4 (2-3a)^2H^2\,.
\ea
Whether  or not this Friedman equation passes cosmological tests is an open question but this merely comes to illustrate how different couplings to matter can lead to exact FLRW solutions and considering slightly more general couplings could have profound consequences for the cosmology.

This type of coupling to matter was shown to be fully free of the BD ghost in Ref.~\cite{Yamashita:2014fga} (beyond the mini-superspace and decoupling limit approximations presented here). Yet we see that the theory admits exact FLRW solutions. This is another example where the existence of exact FLRW solutions in massive gravity clearly does not rely on the existence of a BD ghost.

\section{Summary}
\label{sec:discussion}

While a massless spin-2 field can be perfectly well-defined as a linear theory, consistent coupling with matter requires an infinite set of interactions and leads to the theory of GR.  The same is true for theories of modified gravity where the couplings of the metric(s) to the matter fields have to be constructed in a consistent way. This requirement is necessary both at the classical and the quantum level.

Both the formulation of theories of massive gravity and bi-gravity requires the introduction of two metrics. There is therefore an ambiguity on how matter fields couple to each metrics. In massive gravity, it is usually standard to couple matter covariantly to the dynamical metric.  This choice is stable against quantum corrections and is free of BD ghost. In bi-gravity on the other hand more generic couplings have been considered in the literature. In this work we have explored different couplings of matter to either or both metrics in massive (bi-)gravity.

When matter couples covariantly only to one metric, or when there are two matter fields each coupled to one of the metrics, there is no BD ghost at the classical level. Moreover, quantum corrections from virtual matter fields lead to a cosmological constant for the respective metrics and therefore do not detune the graviton mass and do not lead to any BD ghost. This result can be straightforwardly generalised to multi-gravity theories with each metric coupled to its own matter sector.

In bi-gravity both metrics are put on equal footing and  it may therefore appear quite natural to consider the case in which some matter fields directly couple covariantly to both metrics simultaneously. However, as soon the same matter field couples covariantly to both metrics in the form $\L_\chi=\L_g(g\mn, \chi)+\L_{f}(f\mn, \chi)$, with a kinetic term for $\chi$ in both Lagrangians, the BD ghost appears already classically  at an unacceptably low scale. This was shown explicitly in the mini-superspace approximation as well as in the decoupling limit. Moreover, at the quantum level quantum corrections detune the potential structure at unacceptable low scales.

Motivated by the requirement of maintaining the ghost-free potential structure at the classical and quantum level (at least below a given scale), we have  proposed a new hybrid effective metric to which some matter fields may couple covariantly $\L_{\rm total\ matter}=\L_{\rm eff}+\L_g+\L_f$ with $\L_{\rm eff}=\L(\gd, \chi)$, $\L_{g}=\L(g, \psi_g)$ and $\L_f=\L(f, \psi_f)$ ($\chi$ and $\psi_{g,f}$ representing different independent species). Such couplings $\L(\gd, \chi)$ to the hybrid effective metric are free of the BD ghost in the mini-superspace and in the decoupling limit. On arbitrary backgrounds, we do expect the BD ghost to reappear. However, the absence of ghost in the decoupling limit means that the scale associated with that ghost is larger than the strong coupling scale $\Lambda=(m^2\mpl)^{1/3}$ of the theory. The strong coupling scale can thus be lower that the cut-off of the theory (which can be associated with the mass of the ghost) and there is therefore a non-trivial regime of interest for the theory where the Vainshtein mechanism can still be studied.

In massive gravity with a Minkowski (spatially flat or closed) reference metric, the constraint that removes the BD ghost is also preventing the existence of exact FLRW solutions\footnote{We emphasise, however, that despite the absence of exact FLRW solutions at all scales  in massive gravity, in the scales of the observable Universe, the geometry can still be arbitrarily close to FLRW \cite{DAmico:2011jj}.} \cite{DAmico:2011jj}. This no-go is violated when considering more generic reference metrics (open Minkowski \cite{Gumrukcuoglu:2011ew} or FLRW \cite{Fasiello:2012rw}), or when generalising massive gravity to its quasi-dilaton   or bi-gravity extensions  (see Refs.~\cite{DAmico:2012zv,DAmico:2013kya,Gabadadze:2014kaa,DeFelice:2013tsa,DeFelice:2013dua,Haghani:2013eya} and Refs.~\cite{Akrami:2012vf,Fasiello:2013woa,Comelli:2014bqa,Konnig:2013gxa,DeFelice:2013nba,DeFelice:2014nja,Solomon:2014dua,Schmidt-May:2014iea,Konnig:2014xva}).
With the new coupling to the hybrid effective metric, it is natural to reexplore this no-go and we have shown the existence of exact FLRW solutions in massive gravity with a flat Minkowski reference metric. We also prove that this solution has no BD ghost.
For concreteness and simplicity, this was shown explicitly when a scalar field couples to the effective metric, however the result is much more general than that. In appendix~\ref{app:appendix}, we have considered generic species coupled to the effective metric. For simplicity we focused on scalar field descriptions, but it is easy to see that the results will hold  for other types of fields.
It is important to point out however that the fields that couple to the effective metric do not contribute to the Friedmann equation. As a result most of the known matter fields should still couple to normal  metric $g\mn$ so as to lead to an acceptable evolution of the Universe as was considered in this manuscript.

This finding is encouraging for a number of reasons. First, it opens the venue to a richer class of solutions with cosmological applications in these theories. Second, there are several cosmological tests which can be used to impose constraints on this coupling (such as CMB distortions). We leave these interesting phenomenological questions for future work.

A perturbed ADM analysis however does reveal the existence of a BD ghost at sixth order in perturbations (in an operator which involves the spatial derivatives).  So we do expect the BD ghost to reappear for highly anisotropic solutions of the theory. The implications of this ghost should be worked out carefully to determine the viability of this coupling to matter. In particular the precise cut-off of the theory or mass of the ghost should be established.

Right before submitting this manuscript, an interesting and complementary analysis was posted on the arXiv by Yamashita, de Felice and Tanaka \cite{Yamashita:2014fga}.  The focus of both analysis emphasises different points, our current work has been directed towards the scale of the ghost and the derivation of an effective coupling \`a la $\L_{\rm eff}$ for which the ghost would not appear at or below the strong coupling scale. Ref.~\cite{Yamashita:2014fga} on the other hand has focused on classical considerations and  on couplings  \`a la $\L_\chi$ which are fully free of BD ghost at all scales classically. The techniques involved are thus different but the results are complementary and in full agreement.

\acknowledgments

	We thank Luke Keltner, Andrew Matas, Angnis Schmidt-May, Adam Solomon and especially Gregory Gabadadze,  David Pirtskhalava and Andrew J.~Tolley for useful discussions.
	CdR and RHR are
	supported by a Department of Energy grant DE-SC0009946.
	LH is supported by the Swiss National Science Foundation.
	RHR would like to thank the Perimeter Institute for
	Theoretical Physics (Canada)
	for hospitality and support whilst this work was in progress, and
	DAMTP at the University of Cambridge (U.K.) for hospitality
	near completion of this paper.
	The tensor algebra was performed with the help of
	the xAct package for Mathematica~\cite{xAct}.

\appendix

\section{FLRW solutions with coupling to the effective metric}

\label{app:appendix}

In this appendix, we go beyond the simplest scalar field description of section \ref{sec:FRW} and consider a more general coupling to the effective metric. We start with a generic $P(X,\chi)$ model where $X=-(\p \chi)^2$ and then move to multi-field considerations. The results derived in what follows are not specific to the species being scalar fields, (one could consider for instance vector fields or fermions coupled to the effective vielbein). However when it comes to study homogeneous and isotropic configurations, one would need to deal for instance with the average values of the vector fields. This is straightforward to work out explicitly but beyond the scope of this appendix which is merely to remark on the possibility of having exact  FLRW.

Perfect fluids are an effective description of generic matter fields. For instance if one considered a scalar field with Lagrangian going as $P(X,\chi)$, its associated stress-energy tensor could be described by that of a perfect fluid with energy density $\rho=2XP_X(X,\chi)-P(X,\chi)$ and pressure $p=P(X,\chi)$, (where $P_X(X,\chi)$ represents the derivative of $P$ with respect to $X$).

In GR, or in massive (bi-)gravity where all the matter fields couple to a single metric, one can make the further assumption that the pressure and energy density are only functions of the scale factor in the FLRW case. It is important however to stress that this is an approximation which is an appropriate one in the case of GR (and in the case of matter fields that only couple to one metric), but this approximation is not always an appropriate description of the physics of the system. This is for instance the case for fields that couple to the effective metric since the Lagrangian then depends non-trivially on the \stu fields (through the effective metric). As emphasized earlier  the dependence on the \stu fields is the key behind the violation of the no-FLRW solution. Thus if one were to assume that the fields that couple to the effective metric could be described by an effective fluid whose energy density and pressure only depended on the scale factor, one would of course go back to the original no-go theorem presented in \cite{DAmico:2011jj}. This is of course a trivial statement, which however misses what physically goes on when matter fields couple to the effective metric.

\subsection{Single specie coupled to the effective metric}

We start by considering a single species $\chi$ coupled to the effective metric governed by the Lagrangian $P(X,\chi)$. In addition one may also include other species that couple to the dynamical metric $g\mn$. As mentioned previously, the \stu fields do not enter the Lagrangian of the latter and they can thus be well-described using a perfect fluid approximation where the energy density is only a function of the scale factor.

In order to make the derivation as explicit as possible we introduce the \stu fields. Assuming an exact homogeneous and isotropic symmetry, the dynamical, reference and effective metrics take the form
\ba
\d s_g^2 &=& - N^2(t)\d t^2 +a^2(t)\d x^2\\
\d s_f^2 &=& - \dot \phi^2(t)\d t^2 +\d x^2\\
\label{eq:geff FLRW}
\d s_{\rm eff}^2 &=& - \tilde N^2(t)\d t^2 +\tilde  a^2(t)\d x^2
\ea
with $\tilde N=\alpha N+\beta \dot \phi$ and $ \tilde a = \alpha a+\beta$.
With these considerations in mind, the Lagrangian is then
\ba
\label{Ltotal555}
\L=- 3 \mpl^2 \frac{a \dot a^2}{N}
- a^3 N  \rho_{\rm tot}(a) - \dot \phi W(a)  + \tilde a^3 \tilde N P\(\frac{\dot \chi^2}{\tilde N^2}, \chi\)\,,
\ea
where the contributions from the mass term of the graviton are included in the two functions $\rho_{\rm tot},W$ of $a$ and the energy density of the species that couple directly to $g\mn$ are included in the function $\rho_{\rm tot}(a)$ ($\rho_{\rm tot}(a)$ is expressed in this way as it includes the energy density of the matter fields that couple to the dynamical metric $g\mn$ as well as that which arises from the graviton mass. Notice that $\rho_{\rm tot}$ does not vanish even if no matter fields couple to $g\mn$ and if the cosmological constant and the tadpole vanish $\alpha_0=\alpha_1=0$ since the graviton mass terms ($\alpha_{2,3,4}$) also contribute to it).

Varying with respect to the \stu field $\phi$ leads to the constraint
\ba
\label{eq:cons1}
\frac{\d }{\d t}\left[W(a)-\beta \tilde a^3 (P-2 X P_{,X}) \right]=0\,.
\ea
In the absence of coupling to the effective metric (for instance when $P=0$), we would recover the standard constraint of massive gravity $\p_t W(a)=0$ which prevents the existence of exact FLRW solutions. In this case, this rather provides a constraint for $(P-2 X P_{,X})$ in terms of $W(a)$. This constraint can be read irrespectively as a constraint for $P$, for $P_{X}$,  for $X$, for $\dot \chi$, for $\chi$ or for the lapse $N$. It does matter for which precise variable one decides to solve this constraint, the only relevant aspect is that it stops being a constraint for the scale factor.

The consistency of this constraint equation \eqref{eq:cons1} with the equation of motion with respect to $\chi$ leads to the additional constraint
\ba
\label{eq:Cons2}
H \left[
W'(a)-3 \alpha \beta \tilde a^2 P
\right]=0\,.
\ea
The two constraints can thus be solved for instance as follows
\ba
\label{eq:consts}
P(X,\chi)= \frac{W'(a)}{3\alpha \beta \tilde a^2} \qquad{\rm and} \qquad
X P_X(X,\chi) = -\frac 1{2 \tilde a^3 \beta }\left[W(a)-\frac{\tilde a W'(a)}{3\alpha}\right]\,,
\ea
which could be read as the equations which determine the time-evolution of the field $\chi$ and the \stu field $\phi$ (or the lapse $N$) in time once the evolution of the scale factor is known (which is of course determined by the Friedmann equation).

We see that there is no problem solving  for the constraints \eqref{eq:consts} (apart from very special cases where for instance $P(X,\chi)=Q(X)$ as pointed out in \cite{Solomon:2014iwa}.) However this does not mean that the field necessarily needs to have a potential as was considered for simplicity in section \ref{sec:FRW}. For instance the case where $P(X,\chi)=(1+\chi^2)X$ would also work. As another example we can cite the DBI model \cite{Silverstein:2003hf,deRham:2010eu} which enjoys a shift-symmetry $P(X,\chi)=-f(\chi)^{-1}\sqrt{1-X f(\chi)}$. However we emphasize one does not need to pick a specific form for $P(X,\chi)$. Most choices of functions would be acceptable   (of course as long as $P_X\ne 0$ which is simply asking for the field to actually have a kinetic term), and in the one-field case as long as $P_\chi\ne 0$.

With these expressions for $P$ and $P_X$, the field $\chi$ is of course very constrained and behaves as a perfect fluid with effective equation of state parameter
\ba
\omega_{\rm eff}=-\frac{1}{3\alpha}\frac{\tilde a W'(a)}{W(a)}\,.
\ea
At late--time, we would expect $\tilde a W'(a)/a \to $ const, and the effective equation of state parameter would thus tend to a specific constant which is independent of $\alpha$ and $\beta$.

The resulting Friedmann equation can be derived by varying the Lagrangian with respect to the lapse $N$ and substituting the constraints, leading to the remarkably simple Friedmann equation
\ba
\label{Fried_final}
3 \mpl^2 H^2=\rho_{\rm tot}(a)-\frac{\alpha}{\beta a^3}W(a)\,,
\ea
where once again only contributions from the mass terms enter in $W(a)$, while both contributions from the mass term and from the energy density of the matter fields that couple to $g\mn$ enter in $\rho_{\rm tot}$.

As was pointed out in section~\ref{sec:FRW}, the matter fields that couple to the effective metric do not contribute in the Friedmann equation in the standard way (here we actually see that they do not contribute at all) reason why we should not couple all the fields to the effective metric $\gu\mn$,
most matter fields should  really couple to the dynamical metric $g\mn$.

Even in the absence of a cosmological constant $\alpha_{0,1}=0$ and in the absence of matter fields coupled to the metric $g\mn$, the right hand side of this Friedmann equation does not cancel and includes terms which can lead to a late-time acceleration.

\subsection{Multiple species coupled to the effective metric}

We now generalize the previous derivation and consider several species (symbolized by $\chi_1, \cdots \chi_n$) coupled to the effective metric, so the Lagrangian then takes the form
\ba
\label{Ltotal555}
\L=- 3 \mpl^2 \frac{a \dot a^2}{N}
- a^3 N  \rho_{\rm tot}(a) - \dot \phi W(a)  + \tilde a^3 \tilde N Q\(Y_1, \cdots , Y_n, \chi_1, \cdots, \chi_n\)\,,
\ea
with $Y_j=\dot \chi_j/\tilde N$.

The constraint dictated by the \stu field is then
\ba
\label{eq:Cons1_2}
\frac{\d }{\d t}\left[W(a)-\beta \tilde a^3 \(Q- Y_j Q_{,j}\)\right]=0\,,
\ea
where we use the notation $Q_{,j}\equiv \p Q/\p Y_j$ and the sum over $j=1,\ldots,n$ is implicit. The consistency of this constraint with the equations of motion for the fields $\chi$ leads to the very same second constraint as in \eqref{eq:Cons2}, namely
\ba
\label{eq:Cons2_2}
H \left[W'(a)-3 \alpha \beta \tilde a^2 Q\right]=0\,.
\ea
Now as soon as one has more than one field $\chi$ coupled to the effective metric, one can easily solve both constraints \eqref{eq:Cons1_2} and \eqref{eq:Cons2_2} even if $Q$ did not even depend on the $\chi_j$ (only on the $Y_j$). These two constraints can for instance be seen as constrains for $Y_1$ and $Y_2$ while the other fields $\chi_{j>2}$ satisfy a standard dynamical equation on the effective metric.
However this does not change the fact that these fields do not contribute to the Friedmann equation which remains exactly the same as in \eqref{Fried_final}. This is again the reason why most matter fields should really couple to the dynamical metric $g\mn$, and the species coupled to $\gu\mn$ should really be thought as a dark sector (not dark energy though). However the main point of this interlude of FLRW solutions was to show that as soon as a physical matter field couples to the effective metric (with relatively mild assumptions on that matter), the no-FLRW solution theorem gets violated.

\subsection{Fully generic matter sector}

Finally, we now consider a fully generic sector coupled to the effective metric $\gu\mn$ and a separate sector coupled to the standard metric $g\mn$. As mentioned previously, the sector coupled to $g\mn$ does not {\it directly} interact with the \stu fields and it is sufficient to specify its energy density. As before, we include in $\rho_{\rm tot}(a)$ the energy density of all the species that couple directly to $g\mn$ (and $\rho_{\rm tot}(a)$ also includes contributions that arise from the graviton potential $\sum_n \alpha_n \mathcal{U}_n$.) Keeping the sector coupled to the effective metric  $\gu\mn$ fully general, we have
\ba
\label{Ltotal66}
\L=- 3 \mpl^2 \frac{a \dot a^2}{N}
- a^3 N  \rho_{\rm tot}(a) - \dot \phi W(a)  + \L_{\rm eff}(\gu, \psi_i)\,,
\ea
where $\psi_i$ designate all the species that couple to $\gu\mn$ rather than $g\mn$.

We now proceed precisely as before. Varying this Lagrangian with respect to the \stu field $\phi$ we obtain the constraint
\ba
\label{eq:Cons1_3}
\frac{\d}{\d t}\left[W(a)-\beta \frac{\delta \L_{\rm eff}}{\delta \tilde N}\right]=0\,,
\ea
where $\delta_{\tilde N} \L_{\rm eff}$ can be expressed in terms of the total stress-energy tensor of the species coupled to $\gu\mn$. Indeed, defining the total effective stress-energy tensor
\ba
\tilde T^{\mu\nu}=\frac{2}{\sqrt{-\gu}}\frac{\delta \L_{\rm eff}}{\delta \gu\mn}\,,
\ea
we have (assuming a homogeneous and isotropic metric of the form \eqref{eq:geff FLRW})
\ba
\frac{\delta \L_{\rm eff}}{\delta \tilde N}=\tilde a^3 \tilde T^0_{\ 0}\left[\tilde a, \tilde N, \psi_i,\dot \psi_i\right]\,,
\ea
where we emphasize that for arbitrary particles coupled to $\gu\mn$,  the stress-energy tensor depends non-trivially on $\tilde a$, $\tilde N$ and the fields $\psi_i$ as well as their time-derivatives $\dot \psi_i$ as was explicit in the cases presented earlier.

The equation of motion for the fields coupled to $\gu\mn$ implies the conservation of energy
\ba
D^{(\rm eff)}_{\mu} \tilde T\mupn=0 \qquad \Rightarrow \qquad \p_t \tilde T^0_{\ 0}=-3 \alpha \frac{\dot a}{\tilde a}\(\tilde T^i_{\ i}-\tilde T^0_{\ 0}\)\,,
\ea
where $D^{(\rm eff)}$ designates the covariant derivative with respect to the effective metric and the  expression for $\p_t \tilde T^0_{\ 0}$ is derived assuming homogeneity and isotropy.

Using this expression in \eqref{eq:Cons1_3}, we obtain the two constraints
\ba
\label{eq:Cons1_4}
\tilde T^0_{\ 0}\left[\tilde a, \tilde N, \psi_i,\dot \psi_i\right] &=&\frac{W(a)}{\beta\tilde a^3} \\
\tilde T^i_{\ i}\left[\tilde a, \tilde N, \psi_i,\dot \psi_i\right] &=&\frac{1}{3\alpha\beta\tilde a^3} \(6\alpha W(a)-\tilde aW'(a)\)\,.
\label{eq:Cons2_4}
\ea
These equations are consistent with the results of \cite{Solomon:2014iwa}, however we find here that these constraints can easily be satisfied for an arbitrary sector coupled to $\gu\mn$. In the case of a single field $\psi$ these constraints can for instance be imposed on the field $\psi$ and on the \stu field $\phi$  (which enters in $\tilde N$). In the case of two or more species coupled to $\gu\mn$, there is even more freedom on how these constraints can be interpreted. The end result and the physical observables are however independent of that specific interpretation.

If one assumes a very specific sector for which for instance either  $\tilde T^0_{\ 0}$ or $\tilde T^i_{\ i}$ entirely vanished, these constraints could not be simultaneously satisfied, but this would correspond to a very artificial sector. However we do not know of any physical particle or field for which the resulting $\tilde T^0_{\ 0}$ or $\tilde T^i_{\ i}$ would precisely vanish (one of them may be small compared to the other, which is what is usually considered for dust, but neither contributions to the total stress-energy tensor ever precisely cancel for any physical field). The key element to remember here is that for any physical system neither $\tilde T^0_{\ 0}$ nor $\tilde T^i_{\ i}$ nor $\tilde T^i_{\ i}/\tilde T^0_{\ 0}$ are never just functions of $a$. These quantities are always non-trivial functions of the matter fields $\psi_i$, the shift and the lapse, and here more importantly of the \stu fields. This is illustrated in the previous example of the scalar field, but these results are not specific to the existence of a scalar field (any other physical matter field would share similar properties).

Of course if one assumes that the matter Lagrangian for the sector that couples to $\gu\mn$ has a very specific dependence on $\tilde N$, for instance if one assumes that it is purely linear in $\tilde N$, then the two constraints \eqref{eq:Cons1_4} and \eqref{eq:Cons2_4} cannot be simultaneously satisfied (see \cite{Solomon:2014iwa} for further details on that case), however we emphasize once again that we do not know of any physical particle or field which would give rise to such a Lagrangian (we recall that it is the Hamiltonian that should be linear in the lapse not the Lagrangian. So as soon as the matter Lagrangian includes some independent degrees of freedom beyond that present in the gravity sector, linearity of the Hamiltonian in the lapse automatically imposes a non-linearity of the Lagrangian in the lapse).

We also emphasize that in this derivation we are free to fix the standard lapse to $N=1$, and none of the previous results would differ (although it is now clear that the dependence of the lapse $N$ in $\tilde T^0_{\ 0}$ and $\tilde T^i_{\ i}$ is irrelevant.) Here we only kept the lapse so as to be able to derive the Friedmann equation more easily. Indeed varying the Lagrangian with respect to the lapse leads to the same Friedmann equation found previously
\ba
\label{Fried_final_2}
3 \mpl^2 H^2=\rho_{\rm tot}(a)-\frac{\alpha}{\beta a^3}W(a)\,,
\ea
which confirms once again that the species that couple to $\gu\mn$ do not lead to the standard evolution of the Universe, reason why we have included a separate sector which couples directly to $g\mn$ in section~\ref{sec:FRW} and in this derivation which is encapsulated in $\rho_{\rm tot}$.

	\bibliographystyle{JHEPmodplain}
	\bibliography{references}

\end{document}